\documentclass[aip,numerical,preprint,showkeys,longbibliography,floatfix,superscriptaddress]{revtex4-1}
\usepackage[usenames,dvipsnames]{xcolor}
\usepackage{amsmath}
\usepackage{svg}
\usepackage{subcaption}
\usepackage{natbib}
\newcommand{\Fig}[1]{Fig.~\ref{fig:#1}}
\newcommand{\Tab}[1]{Table~\ref{tab:#1}}
\newcommand{\Sec}[1]{Section~\ref{sec:#1}}

\newcommand{\Eqn}[1]{Eq.~(\ref{eqn:#1})}

\newcommand{\code}[1]{\texttt{#1}}

\newlength{\wholefigwidth}
\newlength{\smallfigwidth}
\newlength{\halfsmallfigwidth}
\newlength{\figwidth}
\begin{document}

\title{ColabFit Exchange: open-access datasets for data-driven interatomic potentials}

\author{Joshua A. Vita}
\altaffiliation{These authors contributed equally}
\affiliation{Department of Materials Science and Engineering, University of Illinois Urbana-Champaign, Urbana, IL 61801, USA}

\author{Eric G. Fuemmeler}
 \altaffiliation{These authors contributed equally}
\affiliation{Department of Aerospace Engineering and Mechanics, University of Minnesota, Minneapolis, MN 55455, USA}

\author{Amit Gupta}
\affiliation{Department of Aerospace Engineering and Mechanics, University of Minnesota, Minneapolis, MN 55455, USA}

\author{Gregory P. Wolfe}
\affiliation{Center for Soft Matter Research, Department of Physics, New York University, New York, NY 10012, USA}

\author{Alexander Quanming Tao}
\affiliation{Department of Aerospace Engineering and Mechanics, University of Minnesota, Minneapolis, MN 55455, USA}

\author{Ryan S. Elliott}
\affiliation{Department of Aerospace Engineering and Mechanics, University of Minnesota, Minneapolis, MN 55455, USA}

\author{Stefano Martiniani}
\affiliation{Center for Soft Matter Research, Department of Physics, New York University, New York, NY 10012, USA}
\affiliation{Simons Center for Computational Physical Chemistry, Department of Chemistry, New York University, New York, NY 10012, USA}
\affiliation{Courant Institute of Mathematical Sciences, New York University, New York, NY 10112, USA}

\author{Ellad B. Tadmor}
\altaffiliation{Corresponding author: tadmor@umn.edu}
\affiliation{Department of Aerospace Engineering and Mechanics, University of Minnesota, Minneapolis, MN 55455, USA}


\begin{abstract}

Data-driven (DD) interatomic potentials (IPs) trained on large collections of first principles calculations are rapidly becoming essential tools in the fields of computational materials science and chemistry for performing atomic-scale simulations.
Despite this, apart from a few notable exceptions, there is a distinct lack of well-organized, public datasets in common formats available for use with IP development.
This deficiency precludes the research community from implementing widespread benchmarking, which is essential for gaining insight into model performance and transferability, and also limits the development of more general, or even universal, IPs.
To address this issue, we introduce the ColabFit Exchange, the first database providing open access to a large collection of systematically organized datasets from multiple domains that is especially designed for IP development.
The ColabFit Exchange is publicly available at \url{https://colabfit.org}, providing a web-based interface for exploring, downloading, and contributing datasets.
Composed of data collected from the literature or provided by community researchers, the ColabFit Exchange currently (September 2023) consists of 139 datasets spanning nearly 70,000 unique chemistries, and is intended to continuously grow.
In addition to outlining the software framework used for constructing and accessing the ColabFit Exchange, we also provide analyses of the data, quantifying the diversity of the database and proposing metrics for assessing the relative diversity of multiple datasets. 
Finally, we demonstrate an end-to-end IP development pipeline, utilizing datasets from the ColabFit Exchange, fitting tools from the KLIFF software package, and validation tests provided by the OpenKIM framework.

\end{abstract}

\maketitle

\section{Introduction} \label{sec:intro}

Leveraging modern computing infrastructures, high-throughput pipelines for density functional theory (DFT) calculations have been able to produce results for millions of atomic configurations spanning a wide range of chemistries and applications \cite{Jain2011,Armiento2011,Saal2013,Emery2017,Palizhati2019,Wines2023}.
These methods have led to the creation of a number of massive datasets of first principles calculations, such as the Materials Project \cite{Jain2013} and the OpenCatalyst Project \cite{Chanussot2020,Tran2023}, among others \cite{Curtarolo2012,Draxl2018,Jha2019,Choudhary2020}, which have served as critical resources for materials discovery and IP development.
While these repositories have proven extremely useful, there still exist opportunities for continued development and dissemination of datasets specifically tailored to fit the needs of developers of data-driven (DD) interatomic potentials (IPs).
In particular, datasets intended for use with IP development typically include a variety of non-equilibrium atomic configurations or hand-selected structures depending on the target application.
Furthermore, datasets intended for fitting DDIPs are often carefully pruned and refined to enable the models to efficiently learn the physical behaviors relevant for the accurate prediction of a given material property, and to achieve stable simulations.
Conversely, existing databases of quantum mechanical (QM) calculations focus predominantly on stable equilibrium structures relevant to material discovery.
Even in the case of databases that do contain portions of the data that may be suitable for use in DDIP fitting, they are rarely organized in a way that facilitates model benchmarking or targeted analysis of model behavior across chemical compound space.

In addition to the issues of content and structure of existing QM calculation databases, common methods for organizing and distributing DDIP training datasets, such as the use of personal Github repositories \cite{Tran2023,Meng2021,Rohskopf2023,Atwi2022,Gardner2022}, Figshare \cite{Christensen2020,Schreiner2022,Takamoto2022,Guan2022,Ramakrishnan2014} or Zenodo \cite{Lysogorskiy2021,Ying2023,Maldonado2023,Wisesa2023} uploads, or other file sharing methods are inconsistent and not conducive to interpretability and interoperability of the datasets.
Datasets stored in this manner often use custom formats (Extended XYZ, HDF5, VASP OUTCARs, CSV, JSON) depending upon the specific research group that generated them, and  despite government insistence \cite{ostpmemo13, ostpmemo22} typically lack metadata necessary for interpretability and reproducibility of the data (missing units, unspecified DFT settings, undocumented inconsistencies in data structure).
Unfortunately, even this limited approach for sharing data is pursued by only a handful of researchers, with the vast majority of DDIP datasets being entirely inaccessible to the general public or made available through private correspondence ``upon reasonable request'', without always honoring such requests.
The end result is a significant decrease in reproducibility of published results and the effective loss of non-trivial amounts of effort and computational time spent on data generation, inevitably hindering scientific progress.

The notion of a FAIR (findable, accessible, interoperable, and reusable) data framework reflects a growing effort in the materials and chemistry communities to address these issues and foster the open exchange of materials and chemical data \cite{Scheffler2022}.
A FAIR database of datasets designed for DDIP training  would help to facilitate collaboration and drive innovation, but must necessarily address a few key issues in order to succeed. Specifically, it must: 1) define a consistent, efficient, and standardized method for storing the data; 2) enable the organization of the data into meaningful, well-documented groupings; and 3) provide tools for easily accessing and contributing to the database in order to promote community engagement.
In this work, we outline a standard for constructing FAIR databases of first-principles calculations, and use it to construct the ColabFit Exchange, the first database of open-access DDIP training datasets.
We will detail the data structure of the resulting database, summarize its content, and demonstrate the use of tools for identifying and characterizing regions of configurational and compositional space sampled by existing datasets.
By serving as a centralized, standardized, and open-access hub for DDIP datasets, the ColabFit Exchange provides the community with a unique opportunity to begin performing large scale analyses of model performances and dataset qualities that were previously infeasible for most researchers.

\section{Structure} \label{sec:methods}

\begin{figure}[!ht]
    \centering
    \includegraphics[scale=0.75]{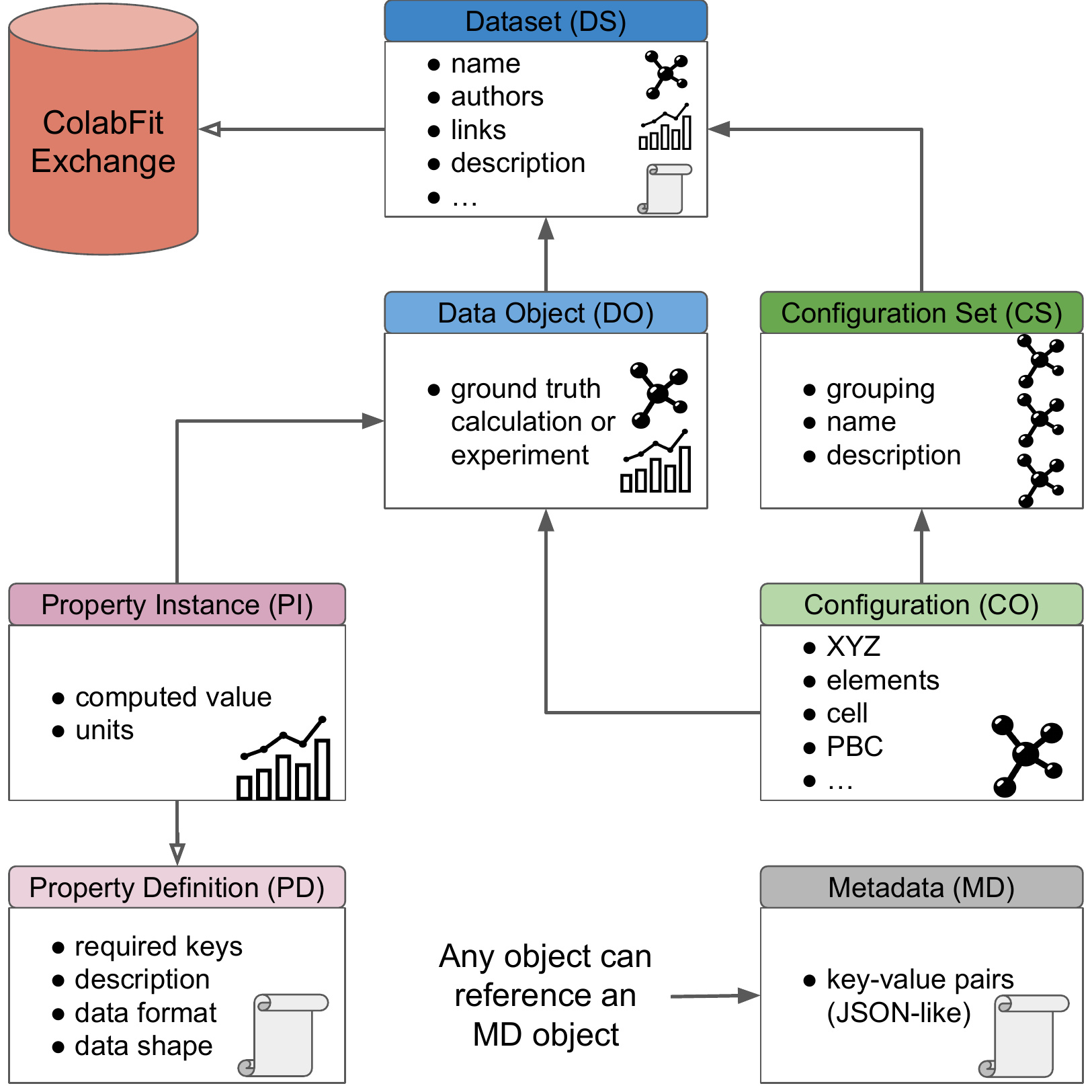}
    \caption{A diagram of the ColabFit Data Standard, which defines the structure of the ColabFit Exchange. The standard comprises seven component types, which can be roughly grouped into three categories (with acronyms defined in the figure): primitive components (PI, CO) for storing input/output data, organizational components (DS, CS, DO) for creating meaningful groupings of lower-level components, and informational components for providing required (PD) or optional (MD) documentation of arbitrary components. Arrows between components specify relationships, e.g., a CO references CS and DO components). Open arrowheads denote many-to-one relationships, while filled arrowheads represent many-to-many relationships. For example, multiple CSs may reference multiple DSs, but each PI references only one PD.}
    \label{fig:standard}
\end{figure}

In order to facilitate the construction of organized datasets, and to ensure that the underlying data is stored in an efficient manner, we develop a hierarchical data storage standard (outlined in \Fig{standard}) comprising seven core components that we describe in detail in this section.
Each of these components is implemented in the \code{colabfit-tools} software package \cite{ColabFit} following an object-oriented design scheme.
In this section we will give examples of how the ColabFit Data Standard can be applied to construct a database of atomistic ground-truth datasets, as this is the primary task which the ColabFit project aims to address. It is important to note, however, that the data standard is designed to be sufficiently flexible for adaptation to many other scientific domains where data-driven approaches are of interest.

\subsection{Low-level components (COs and PIs)}
\label{sec:methods:co_pi}

The two fundamental building blocks of the ColabFit Data Standard are Configurations (COs) and Property Instances (PIs). Each CO stores a representation of an elementary object of interest and typically serves as input ($\mathbf{x}$) to a DD pipeline. PIs, on the other hand, store instances of property measurements associated with COs and typically serve as predictive targets ($\mathbf{y}$). For the examples outlined below, these will be atomistic configurations and target property values measured through ground-truth calculations or through experiments. 

Broadly speaking, a CO subclass must define two critical functionalities: 1) it must define a list of keys whose values are used to generate a hash for comparing CO objects, and 2) it must define two functions, one for generating a dictionary of information summarizing the contents of the CO, and another specifying how information from a \textit{set} of COs may be aggregated into a single dictionary.
These summary and aggregation functions will be called by higher-level objects to gather information about groups of COs.
For example, in the case of an atomic configuration, the atom types, Cartesian coordinates of the atoms, cell vectors, and periodic boundary conditions would all be required to uniquely distinguish between two COs.
A summary dictionary for an atomic configuration could include information such as the number of atoms in the cell, the chemical formula, the periodicity of the cell, or any other information deemed useful by the curators of the dataset.
These traits enable the development of workflow pipelines for aggregating information about groups of configurations up to a higher-level component (see \Sec{methods:do_cs_ds}), which in turn aid in the construction of rich and efficiently queryable metadata.

Notably, the ColabFit Exchange currently makes the assumption that a given database is used to store only one type of CO at a time (e.g., only atomic configurations) in order to simplify the data aggregation process.
This assumption may be relaxed in the future, depending upon the needs of the community.
Using the ColabFit Data Standard to construct a database for data other than atomistic property predictions (e.g., property prediction for biomolecules specified by sequences whose characters span 20 naturally occurring amino acids) will typically involve writing a new CO subclass specification, with required keys matching the application of interest and custom aggregation functions.

Whereas COs store the input, PIs store the ``ground-truth'' output. Importantly, a PI contains a \textit{single} computed property (and its units), such as the potential energy of the system or the atomic forces, rather than all of the properties associated with a given calculation. The decision to separate each property into its own PI allows for more efficient data storage, as it means that duplicate documents do not need to be stored in the database even in the case where two calculations have only a subset of matching properties (e.g., DFT calculations of two different single-atom primitive cells of ground-state crystals, which would both have zero forces, but will likely have different energies).
Furthermore, this design choice allows PIs to be added or modified independently of the corresponding COs, which helps to simplify the process of cleaning and modifying datasets.
In practice, a PI is a dictionary of key--value pairs for storing computed or measured properties and their associated units, plus some basic functionality for unit conversion and hashing.
All PIs are required to point to exactly one Property Definition object in order to properly document the structure and contents of the PI (see \Sec{methods:pd_md} for more details).

\subsection {Informational components (PDs and MDs)}
\label{sec:methods:pd_md}

With the goal of encouraging reproducibility and ensuring that all of the data stored within a ColabFit database is well-documented \cite{ostpmemo13,ostpmemo22}, Property Definition (PDs) and Metadata (MD) objects can be used to enforce structure in the data and provide additional information about each object.

All PIs are required to point to exactly one PD, which serves as an explicit, computer-readable definition (schema) of the contents of the PI following the KIM Property Definition specification \cite{kimprop}.
The most important benefit of PDs is that they improve the homogeneity of the database by ensuring that all properties of the same type are stored in the same format.
PDs specify all of the keys available in the PI; for each of these keys, the PD will also specify if the key is required/optional, the data type of the corresponding value, the shape of the data (i.e., scalar, vector, tensor, ...), if the value has units, and a brief description of the data.
The KIM PD specification also supports uncertainty information for stored values, which may be included in ColabFit in the future.

A simple atomistic property example is the \code{potential-energy} PD, which has the keys ``energy'' (the potential energy of the system; required, float, scalar, has units), ``per-atom'' (if the energy has been divided by the number of atoms in the CO; required, boolean, scalar, unitless), and ``reference-energy'' (the value, if any, which has been subtracted from the ``energy'' value; optional, float, scalar, has units).
As is the case with COs and PIs, by storing the PD as its own object rather than attaching the data directly to each PI, we are able to avoid duplicating data unnecessarily while still maintaining proper documentation of the PI contents.

While PDs serve as mandatory documentation of the contents of a PI, MD objects can be used to store optional additional information about objects of any type. MD objects can be any valid JSON dictionary, and are intended to be sufficiently flexible for storing data that does not fit naturally into any of the other object types.
One of the most common applications of MD objects for constructing a DFT database would be to store pointers to raw input/output files (e.g., INCAR/OUTCAR files from VASP \cite{Kresse1993}) or additional information regarding simulation settings. Best practice would be to use MD objects to ensure that sufficient information is provided to reproduce any calculation in the database.
In addition to improving reproducibility, proper use of MDs can also be valuable for identifying when datasets were computed using different settings or levels of theory, which can be important for transfer learning tasks \cite{Hutchinson2017,Jha2019} and can inform on when datasets may, or may not, be used in conjuction with each other for model training.
Generally, the contents of MDs are not expected to be queryable, as available keys may vary drastically between datasets, though in some cases we found it useful to manually parse the MDs to improve the quality of common queries over COs or PIs (e.g., descriptive labels on COs, or levels of theory used for computing PIs).

\subsection{Organizational components (DOs, CSs, and DSs)}
\label{sec:methods:do_cs_ds}

Given that the ColabFit Data Standard is meant for constructing databases for data-driven model development, it obviously must allow for the data to be organized in meaningful and useful ways. Data Objects (DOs), Configuration Sets (CSs), and Datasets (DSs) facilitate this by defining higher-level groupings of lower-level objects.

A DO is perhaps the simplest of these groupings---it defines relationships between one or more COs with one or more PIs. Conceptually, DOs should be used to link inputs and outputs of a given calculation or measurement.
For example, a DFT calculation would typically produce both an energy PI and an atomic forces PI, which could be grouped under a single DO that also points to the corresponding CO and details of the calculation in an MD.
A more complex example would be a nudged elastic band calculation \cite{JNSSON1998}, where it would be necessary to define a relationship between a computed energy barrier (a PI) and multiple images interpolating between the start/end transition states (each stored as their own CO).

Another object, which we observe is particularly useful in practice for improving data interpretability, is the CS. A CS defines a grouping (and optionally, an ordering) over one or many COs, and allows a user to give a name and a description to that grouping. Generally, CSs should be used for organizing configurations into groups that will help end-users better understand the contents of the dataset. In the materials and chemical sciences, it is common for dataset developers to organize their data based on attributes such as molecule type, physical structure, or method of generation \cite{Chmiela2017,Wen2020,Smith2021}.
For example, molecular dynamics or relaxation trajectories are often grouped together by DDIP developers.
Similar methods can be useful in other deep learning fields, such as with the MNIST \cite{deng2012mnist} or CIFAR-10 \cite{Krizhevsky09learningmultiple} datasets where the data are naturally grouped by class.
Such groupings make it easier for users of the datasets to understand the contents of the dataset, facilitate filtering, and improve interpretability of the behaviors of models trained to the data.

The highest level object (aside from a database itself) is a DS, which matches the canonical meaning of the word: a collection of data points and any associated metadata.
Similar to how a CS defines a collection of COs, a DS defines a collection of DOs and CSs, and includes additional metadata such as a name, list of authors, relevant links, and a description.
Notably, a DS references CSs rather than COs directly in order to ensure that any organizational structure imposed by the CSs is reflected in the DS as well. The DS serves as a complete, well-documented, and queryable representation of a collection of computed values and their corresponding inputs, and is intended to be packaged and distributed as a self-contained object to facilitate reproducibility, standardized benchmarking, and collaboration.
All DSs currently in the ColabFit Exchange are assigned unique DOIs for tracking citations and can be downloaded at \url{https://colabfit.org} as extended XYZ files in a standardized format.

\subsection{Additional technical details}

Two important features of the ColabFit Data Standard are the abilities to store the data in an efficient and queryable manner, and to aggregate low-level information in order to generate information-rich, high-level metadata. While part of this functionality is achieved through careful separation of data objects into their constituent parts (PIs, COs, PDs, and MDs), it also depends upon a few other technical details discussed in this section.

First, hashing functions are used  to generate unique IDs for every component in the database; these digest specified contents of each component and return a hexadecimal string.
The contents of a component that are digested in order to generate the hash vary depending on the component's type: MDs directly hash their entire contents; PIs hash their computed values and units; COs hash the contents of their required keys.
The hashes for higher level components (DOs, CSs, and DSs) are generated by hashing the IDs of all of their sub-components. For example, a CS's ID is a hash of the list of the IDs of all COs grouped by the CS.
PDs are the only components which do not use hashes for their unique ID, but instead are given user-specified names, as there are relatively few PDs and it is important for their IDs to be human-readable.
This hashing avoids the issue of duplicate entries (those whose content is identical within machine precision) when users re-upload portions of existing datasets or coincidentally generate the exact same data as another author (a relatively common occurrence in the materials and chemical sciences).

Second, aggregation pipelines were developed for building metadata for high-level objects (CSs and DSs).
Although some metadata is stored on CS/DS objects directly, other information must necessarily be propagated up from the CO/PI level; for example, information such as the total number of atoms contained within a CS, the chemical formulas present, or the relative concentrations of elements.
In order to enable this type of data aggregation, low-level components (COs and PIs) provide functions for returning ``summaries'' of their contents, which are key--value dictionaries summarizing any additional information of interest that the database authors think might be useful. The low-level components also provide functions for merging lists of metadata dictionaries into a single dictionary.
Database developers may adjust the behaviors of these summary and aggregation functions depending on their needs and target applications.
This aggregated metadata greatly improves the queryability and interpretability of the data, and helps to build a database that can be more easily used by model developers for drawing insights about their data.

\subsection{Comparison to OPTIMADE}

In order to simplify the process of understanding the design choices made in this work, we compare the ColabFit Data Standard outlined above to the OPTIMADE API \cite{Andersen2021}, which is a broad effort from researchers across many domains of materials science to develop interoperable databases of materials data.
Although the ColabFit Exchange is not yet OPTIMADE-compliant (which is a future goal of the work), many parallels can be drawn between the components described in \Fig{standard} and objects from the OPTIMADE API.
Given the ubiquity within the community of the need for representing atomic configurations, it is unsurprising that the CO object described in \Sec{methods:co_pi} contains all of the information necessary to define a \texttt{Structures} object in the OPTIMADE API, and could be easily made to match with some additional processing (i.e., storing various chemical formulas, or re-formatting certain fields to fit the OPTIMADE specifications).
The ColabFit Standard PD and PI components roughly correspond to OPTIMADE \texttt{Property Definition} and \texttt{Calculation} objects, though the two standards begin to diverge in the specific details of these components.
For example, PDs allow for specifying units, whereas the OPTIMADE \texttt{Property Definition} does not, and PIs are required to be associated with a PD while OPTIMADE \texttt{Calculation} objects are not.
The largest discrepancies between the two standards arise from the higher level components described in \Sec{methods:do_cs_ds}, where ColabFit's need for defining groupings over objects (e.g., CSs as groups of COs, and DSs as groups of CSs and DOs) are not well-supported by the current OPTIMADE API.
Although possible workarounds exist in order to represent a DO/CS/DS using existing OPTIMADE objects (e.g., with relationships), such constructions would have been inefficient and lacking in many of the desired functionalities of DOs/CS/DSs.
There is, however, a current effort within the OPTIMADE community to support trajectory-like objects (groups of \texttt{Structures}, intended for storing simulation trajectories) which, once fully implemented, will more easily support the needs of the ColabFit Exchange.

\section{Overview}

\Tab{objects} provides a summary of the contents of the ColabFit Exchange, which is currently (September 2023) composed of 139 unique datasets contributed by their authors or gathered from the literature.
\begin{table}[!ht]
    \centering
    \caption{Counts of objects of interest in the ColabFit Exchange, excluding the data from the OpenCatalyst datasets. These values do \textit{not} double count in the case where there exist duplicates of a given object (e.g., when an identical configuration was uploaded in multiple datasets, or an author is credited on multiple publications). Here, a ``chemical system'' refers to a set of unique constituent atom types.}
    \begin{center}
        \begin{tabular}{lr}
            \hline\hline
            Objects &  Count \\
            \hline
            Datasets & 139 \\
            Configuration sets & 459 \\
            Data objects & 11,185,734 \\
            Configurations & 10,752,923\\
            Atoms & 512,108,838 \\
            Chemical systems & 68,474 \\
            Publications & 79 \\
            Authors & 323 \\
             \hline\hline
        \end{tabular}
    \end{center}
    \label{tab:objects}
\end{table}
These datasets are further broken down into 459 configuration sets, which can be readily combined, split, or grouped in order to define new datasets based on the needs of the community.
In total, the ColabFit Exchange contains over 11 million DOs, corresponding to approximately 28 million computed properties.
Note that the OpenCatalyst datasets (which are included in the ColabFit Exchange) are not included in these summary statistics, as they are already well-documented elsewhere in the literature \cite{Chanussot2020,Tran2023} and their large sizes ($\sim$134 million DOs for OC20) would obscure the results from the other datasets.
As the ColabFit Exchange continues to grow, updated statistics summarizing its contents can be found at \url{https://colabfit.org}.

\begin{figure*}[!ht]
    \centering
    \includegraphics[scale=1.0]{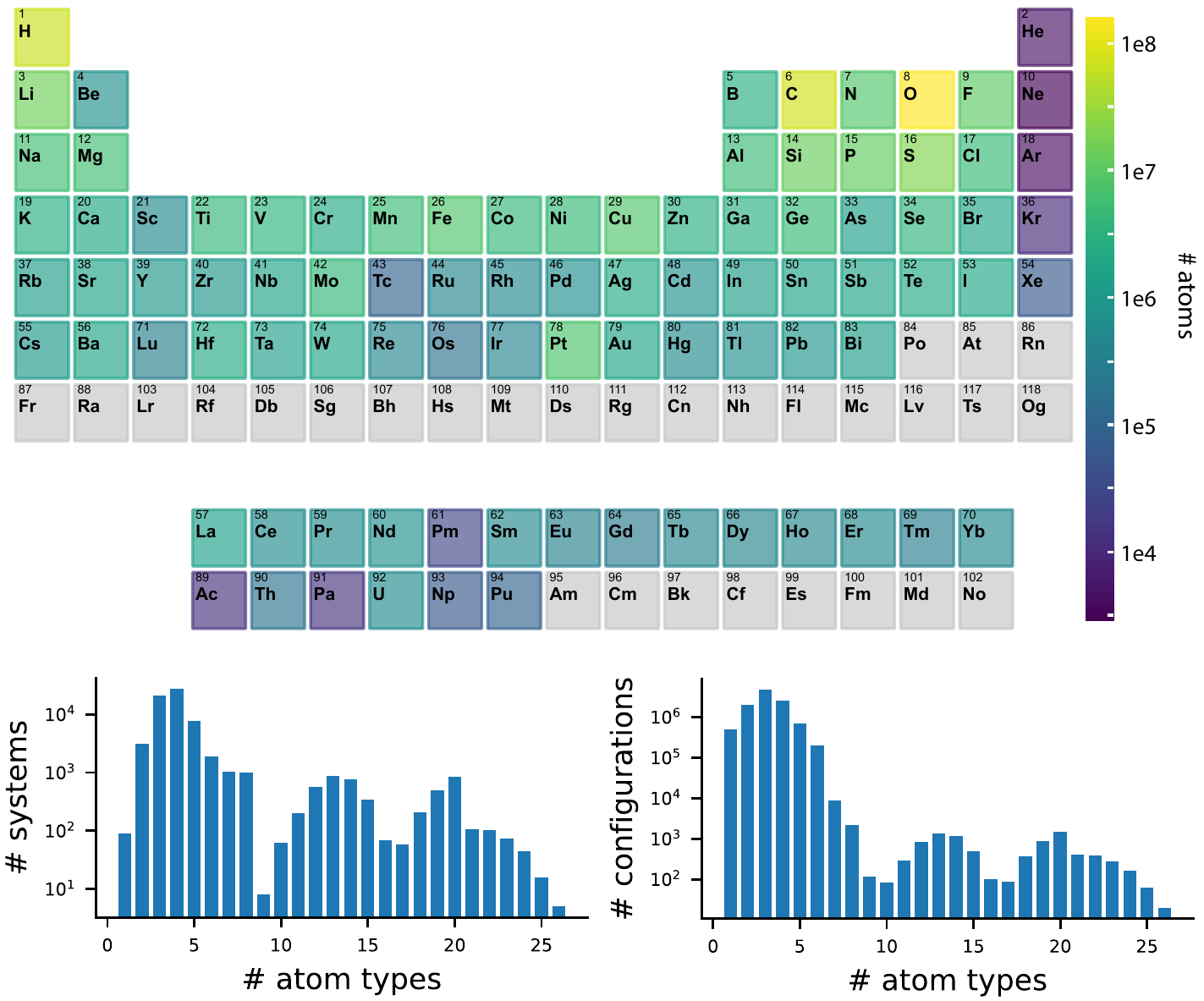}
    \caption{Chemical composition of the ColabFit Exchange, spanning 89 of the 118 elements on the periodic table, for a total of 68,474 unique chemical systems. After excluding the OpenCatalyst data (which is not represented in this figure), the majority of the database is composed of organic molecules (C, H, and O alone make up $\sim$60\% of the data shown in this figure) due to the relative popularity and availability of molecular datasets. There is currently no data for elements with atomic numbers between 84 and 88, or greater than 94. The bottom panel shows histograms of the number of unique chemical systems (left) or configurations (right) present in the ColabFit Exchange for different numbers of atomic types (i.e., the number of unary/binary/ternary/... systems or configurations). 
    The HME21 dataset \cite{Takamoto2022} accounts for the majority of the data with large numbers of atom types; without HME21, all systems have fewer than 10 atom types.}
    \label{fig:ptable}
\end{figure*}

\begin{figure}[!ht]
    \centering
    \includegraphics[scale=1.0]{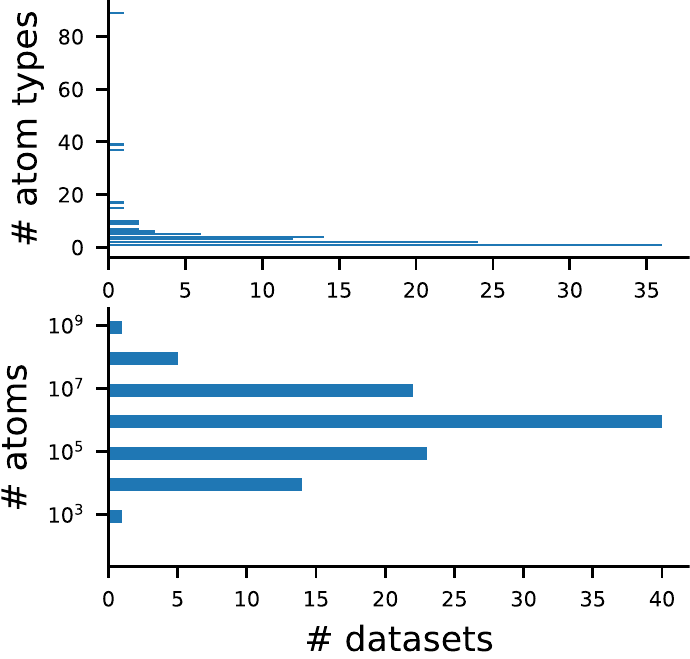}
    \caption{Histogram showing the sizes of the datasets currently in the ColabFit Exchange. The distribution of the total number of atoms summed over all COs in a given dataset is Gaussian-like, centered about a mean of $10^6$.}
    \label{fig:ds_natoms_histograms}
\end{figure}

The $\sim$11 million atomic configurations (for a total of ~512 million atoms) spanning nearly 70,000 chemical systems can be further analyzed based on their chemical composition, as shown in \Fig{ptable}.
Here, a ``chemical system" is defined as a set of unique constituent atom types, e.g., C, C-H, C-H-N, \ldots, and is indicative of the types of chemistries explored within the ColabFit Exchange.
Though single element datasets are the most common (see \Fig{ds_natoms_histograms}), 95\% of the configurations in the ColabFit Exchange include at least two elements, meaning the ColabFit Exchange may be used as a starting point for the development of many multi-element models.
Much of the multi-element data comes from larger datasets designed for the construction of ``universal'' IPs intended to model all relevant types of atomic interactions \cite{Smith2018,Chen2022,Deng2023}, such as the Materials Project trajectory dataset \cite{Deng2023}, and others from the literature \cite{Takamoto2022,Chen2022,Komissarov2022}.
By providing access to all of these datasets within a unified framework, the ColabFit Exchange will simplify the process of constructing training datasets for new chemical systems that have not yet been explicitly sampled by the datasets currently in the ColabFit Exchange.

The values in \Tab{properties} provide a further breakdown of the most prevalent computed properties stored within the ColabFit Exchange that are available for supervised training.
Energies are the most commonly computed property, followed by forces.
Note that the energy counts in \Tab{properties} are a sum over the four types of energy PIs specified by the publications associated with the datasets in the ColabFit Exchange (potential, free, atomization, and formation energy), where each energy type is given its own PD.
Note, the raw number of force PIs shown in \Tab{properties} does not reflect the total number of individual atomic force vectors in the ColabFit Exchange---the number of individual force vectors is much higher, approximately equaling the number of atoms in the database multiplied by the fraction of DOs that contain an atomic force PI (90\%).
Stresses are available for only about half of the DOs in the ColabFit Exchange, with the majority coming from the Materials Project (MP) trajectory dataset \cite{Deng2023}.
The ColabFit Exchange also includes, for subsets of the data,  additional properties that are supported within the framework as their own PDs but are less relevant to DDIP development.
These additional properties include indirect and direct band gaps, magnetization, atomic charges, polarizability, dipole moments, and a large collection of common molecular properties from datasets like those derived from GDB-17 \cite{Ruddigkeit2012}.

\begin{table}[!ht]
    \centering
    \caption{Counts of property instances in the ColabFit Exchange, excluding the data from the OpenCatalyst datasets. These values \textit{do} double count in the case where two identical copies of a property exist (e.g., two distinct configurations were uploaded with identical potential energies) in order to accurately reflect the number of target values in the ColabFit Exchange.  
    Though many of the datasets currently in the ColabFit Exchange contain more computed properties than the three shown here, energies, forces, and stresses are the three that are predominantly used for training DDIPs.}
    \begin{center}
        \begin{tabular}{lr}
            \hline\hline
            Property Instance (PI) &  Count \\
            \hline
            Energy & 11,293,268 \\
            Atomic forces & 10,102,772 \\
            Cauchy stress & 6,729,342 \\[5pt]
            Total & 28,125,382\\
            \hline\hline
        \end{tabular}
    \end{center}
    \label{tab:properties}
\end{table}

At the dataset level, \Fig{ds_natoms_histograms} shows that the ColabFit Exchange has a wide range of dataset sizes, both in terms of the total number of atoms and the number of unique atom types contained within a given dataset.
Though single element datasets are the most common, these datasets are typically smaller than multi-element datasets.
The three datasets with greater than 20 atom types are HME-21 \cite{Takamoto2022}, the Materials Project trajectory dataset \cite{Deng2023}, and the elpasolite crystal dataset \cite{Faber2016}.
The number of molecular datasets versus the number of condensed matter datasets is roughly evenly split (51 molecular, 50 condensed matter, and 5 mixed), though the molecular datasets usually include significantly more atomic configurations due to their smaller number of atoms per configuration.

\section{Applications}
\label{sec:dataset_analysis}

A critical step towards improving DDIP design and efficiently constructing models for specific applications is to gain a better understanding of what regions of composition and configuration space have, or have not, been sampled by existing datasets.
As the ColabFit Exchange is the first attempt at curating an exhaustive list of DDIP-fitting datasets, it provides a unique opportunity for performing this type of analysis.
Towards this end, in this section we explore the use of tools for identifying and characterizing regions of overlap between two datasets.
Furthermore, we demonstrate how the ColabFit Exchange can integrate with other model fitting and validation tools to create an end-to-end fitting framework.

\subsection{Comparing atomic environments}
\label{sec:m3gnet}

\begin{figure}[!ht]
    \centering
        \includegraphics[scale=0.4]{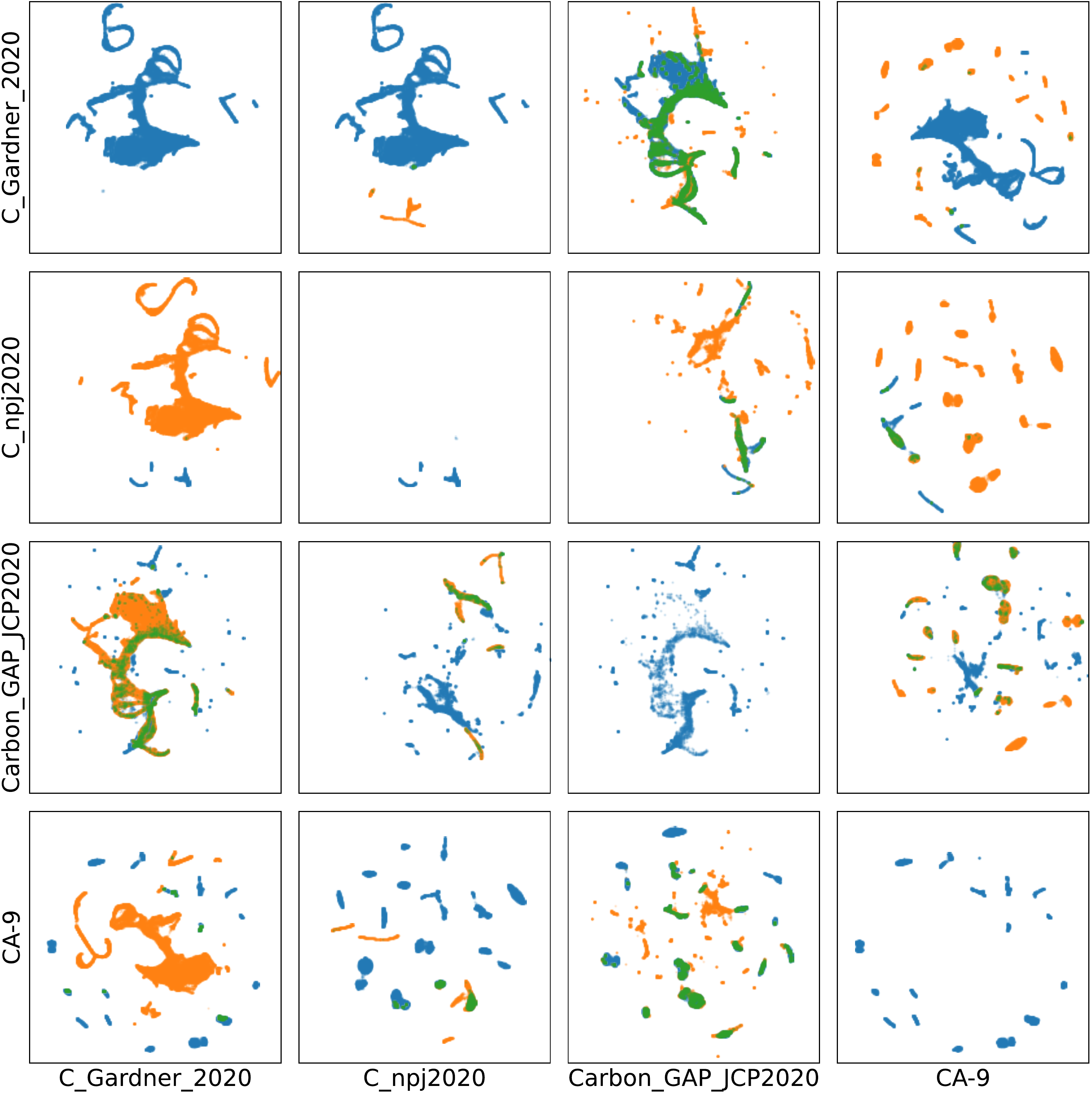}
    \caption{Visualizations of the configurations in the \texttt{C\_Gardner\_2022}, \texttt{C\_npj2020}, \texttt{Carbon\_GAP\_JCP2020}, and \texttt{CA-9} datasets in relation to each other. Plots are generated by applying UMAP to configuration-averaged descriptors extracted from the M3GNet model, as described in \Sec{m3gnet}. Row labels denote the ``reference" dataset used for DCA in \Sec{dca}, which are colored blue in each panel. Column labels denote the ``evaluation" dataset, and are colored orange. To help highlight regions of overlap, points from the reference dataset have been colored green if there is at least one point from the evaluation dataset within a chosen threshold value. Panels along the diagonal correspond to only the reference set, in order to help guide visual comparisons to the other panels in the same row. Note that UMAP embeddings were performed individually for each panel, including only the two datasets within that panel. This means that the embeddings may not be identical even for the same dataset across rows or down columns.}
    \label{fig:dca_umaps}
\end{figure}

In order to compare configurations between datasets, it is convenient to first define a method for obtaining a vector representation of the atomic environments in the configurations (which is invariant to permutations, rotations and translations).
This can be done using several well-documented local ``descriptors,'' such as the Atom-Centered Symmetry Functions (ACSF)~\cite{Behler2011} or Smooth Overlap of Atomic Positions (SOAP)~\cite{Bartok2013} descriptors, among others~\cite{Rupp2012,Drautz2019,Huo2022}.
However, given the quadratic scaling of the sizes of local environment descriptors with the number of atom types, this rapidly becomes intractable when performing database-wide analyses, as is the goal here.
We instead choose the descriptor to be a learned-representation, i.e., intermediate vectors generated by a pre-trained graph-based model.
For this task, we chose to use the M3GNet universal potential \cite{Chen2022}, which has been previously trained to a subset of the Materials Project relaxation trajectory dataset.
The learned representation is taken from the final layer of the M3GNet model prior to the regression head, which has a size of $N_{atom}\times64$, regardless of the number of chemical species in the atomic configuration.
These $N_{atom}\times64$ matrices are then averaged over $N_{atoms}$ in order to produce a single length-64 vector for each atomic configuration.
UMAP visualizations of these configuration-averaged M3GNet representations are shown in \Fig{dca_umaps}.

\subsection{Delaunay Component Analysis (DCA)}
\label{sec:dca}

While visualizations like those shown in \Fig{dca_umaps} are commonly used for obtaining a qualitative understanding of the contents of a dataset, and often provide advantages over methods like PCA, the use of UMAP (or tSNE \cite{Maaten2008}) makes it challenging to obtain quantitative metrics since distances are not preserved between the original and embedded spaces.
In order to obtain a more quantitative understanding of the relationships between datasets, we explore the recently developed Delaunay Component Analysis (DCA) technique \cite{Poklukar2022} to quantify the overlap between two datasets.
Originally intended for comparing between the manifolds of two learned representations of the same data, we instead apply DCA here to the separate, yet related, task of comparing two datasets under the same representation (i.e., the learned M3GNet latent vectors).
Though we provide a brief summary of the DCA method here, for a more thorough explanation we refer the reader to Ref.~\citenum{Poklukar2022}.
Some additional analysis of DCA as it relates to this work can be found in the Supplementary Material.
The DCA analysis shown in this section uses the code provided by Ref.~\citenum{Poklukar2022}, which is included in the \code{colabfit-tools} package alongside a growing set of tools for dataset analysis organized under the \code{colabfit-analyze} sub-package.

The goal of DCA is to derive metrics quantifying the degree of overlap between two manifolds, where one manifold is defined by points in a ``reference" dataset, and the other manifold is defined by points in an ``evaluation" dataset.
In this case, the manifolds exist in the 64-dimensional latent space of the M3GNet model from which we extracted the descriptors, and represent the phase spaces sampled by each dataset.
DCA constructs an approximate Delaunay graph (known as the ``dual graph'' of a Voronoi diagram, where the circumcenters of triangles in the Delaunay graph are the vertices of the corresponding Voronoi diagram) of the manifolds, then distills the graph into connected components, i.e., robust sub-graphs, using a minimum spanning tree.
Vertices in the Delaunay graph correspond to data points from the reference or evaluation datasets; edges link points which are ``natural neighbors" of each other (i.e., they have adjoining Voronoi cells).
Connected components are sub-graphs representing clusters in the representation space, and may be composed of a mix of vertices from both the reference and evaluation datasets.
Note that DCA does not modify the representations of the configurations (descriptors) in any way, so it inherits all attributes of the M3GNet descriptor (e.g., invariance to rotations of configurations, learned embeddings of atomic types, etc.).
Using the distilled components, DCA then evaluates a ``consistency'' ($c$) and ``quality'' ($q$) score for each component, defined as:
\begin{equation}
\label{eqn:dca_cq}
\begin{gathered}
    c(\mathcal{G}_i) = 1 - \frac{|\, |\mathcal{G}_i^R|_\mathcal{V} - |\mathcal{G}_i^{E}|_\mathcal{V} \,|}{|\mathcal{G}_i|_\mathcal{V}}\\
    q(\mathcal{G}_i) = 
    \begin{cases}
    1 -  \frac{(| \mathcal{G}_i^{R}|_\mathcal{E} + |\mathcal{G}_i^{E}|_\mathcal{E})}{|\mathcal{G}_i|_\mathcal{E}} & \text{if } |\mathcal{G}_i|_\mathcal{E} \geq 1,\\
    0              & \text{otherwise},
    \end{cases}
\end{gathered}
\end{equation}
where $\mathcal{G}_i$ is the Delaunay graph of component $i$, and $|\mathcal{G}_i^R|_\mathcal{V}$ and $|\mathcal{G}_i^R|_\mathcal{E}$ denote the cardinalities of the vertex and edge sets of $\mathcal{G}_i$ restricted to dataset $R$, respectively. Conceptually, consistency measures how evenly represented each dataset is within a component, while quality measures how well mixed the datasets are in a component.
The local metrics of consistency and quality, which are computed individually for each component, can then be used to identify ``fundamental'' components (those with both high consistency and high quality) in order to calculate global metrics of
``precision'' $p$ and ``recall'' $r$ between the two datasets, defined as:
\begin{equation}
    p = \frac{|\mathcal{F}^E|_\mathcal{V}}{|\mathcal{G}^E|_\mathcal{V}} 
    \quad\text{ and }\quad
    r = \frac{|\mathcal{F}^R|_\mathcal{V}}{|\mathcal{G}^R|_\mathcal{V}}, \\
    \label{eqn:dca_pr}
\end{equation}
where $\mathcal{F}^E$ and $\mathcal{F}^R$ refers to the sub-graphs of the evaluation and reference datasets, respectively, which are contained within a fundamental component.
Intuitively, the precision measures the fraction of points from the evaluation dataset which overlap with the reference dataset.
Recall measures how well the reference dataset is represented by the evaluation dataset.
A high precision score means that the evaluation dataset is well contained by the reference dataset; a low recall means that the reference dataset includes data which is not well-represented by the evaluation set.
These definitions of precision and recall are similar to those commonly used in other deep learning tasks for quantifying the degree of overlap between two distributions, though the use of ``fundamental components" is a valuable modification unique to DCA which helps apply the metrics to manifold analysis.

As a demonstration of the utility of the global metrics of precision and recall, we perform DCA using four datasets from the ColabFit Exchange which include only pure carbon data: \texttt{C\_Gardner\_2020}~\cite{Gardner2022}, \texttt{C\_npj2020}~\cite{Wen2020C}, \texttt{Carbon\_GAP\_JCP2020}~\cite{Rowe20}, and \texttt{CA-9}~\cite{Hedman2021}.

The \texttt{C\_Gardner\_2020} dataset contains DDIP-computed molecular dynamics trajectories of a melt/quench/anneal process;
\texttt{C\_npj2020} has a relatively narrow focus, with an emphasis on monolayer and bilayer graphene, diamond, and graphite structures; \texttt{Carbon\_GAP\_JCP2020} contains a wide variety of carbon systems, e.g., bulk, liquid, nanotubes, fullerene, graphene, etc.; and, finally, \texttt{CA-9}~\cite{Hedman2021} has DFT-computed molecular dynamics trajectories of nine carbon allotropes (diamond, lonsdaleite, graphene, haeckelite, SWCNT, fullerene, cumulene, carbyne, and amorphous C).
We extract the configuration-averaged M3GNet representations for each dataset, as described in \Sec{m3gnet}, and use these as the representations for DCA.

\begin{figure}[!ht]
    \centering
        \includegraphics[scale=0.75]{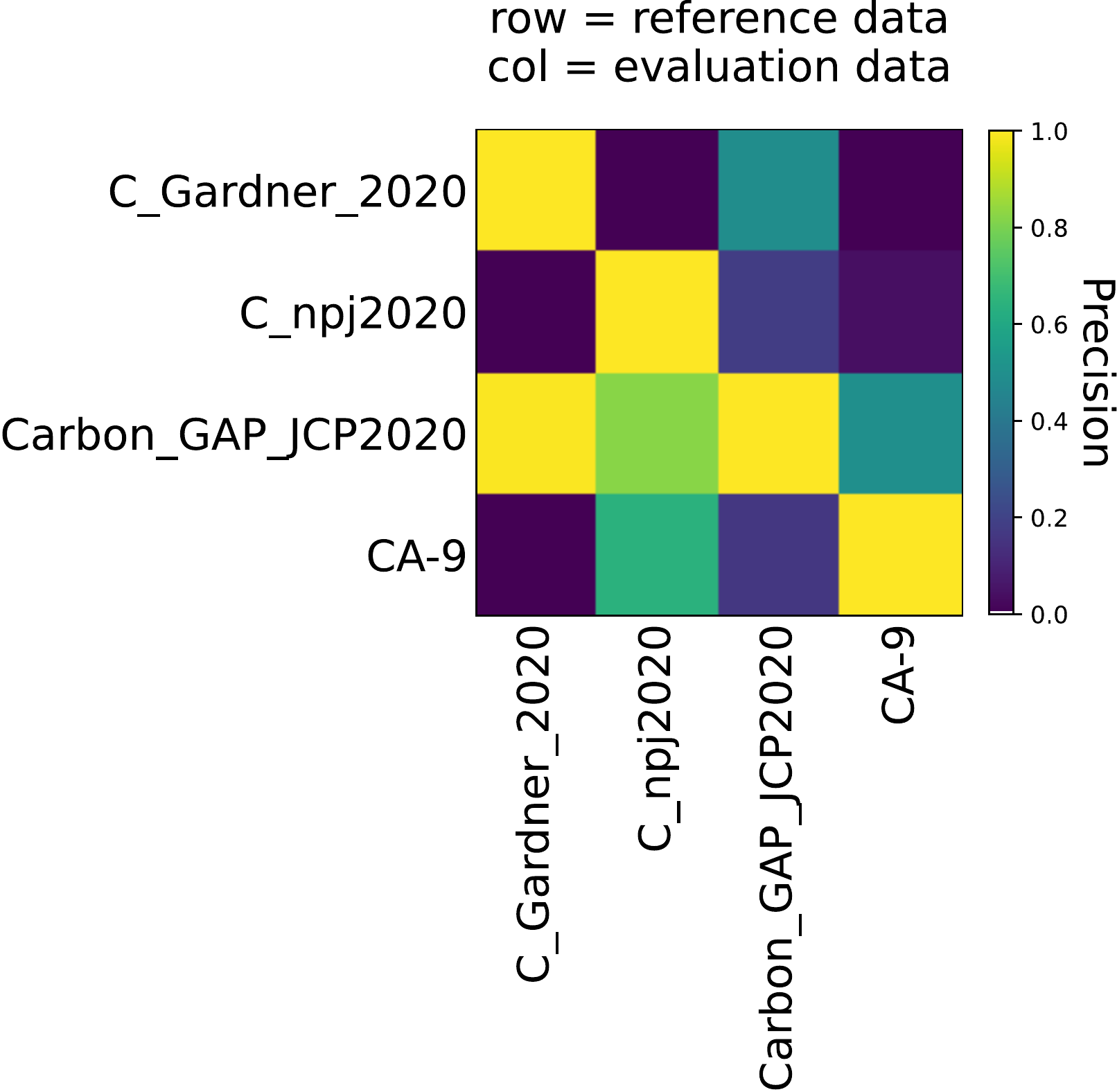}
    \caption{Precision scores obtained by DCA comparing the four datasets from \Fig{dca_umaps} to each other. A high precision score means that the evaluation dataset (column labels) is well-contained within the reference dataset (row labels). A high recall score (which corresponds to the transpose of this matrix) means that the evaluation dataset provides good sampling of all components of the reference dataset.}
    \label{fig:dca_prmat}
\end{figure}

The precision scores reported in \Fig{dca_prmat} immediately provide quantitative insights which match our intuitions based on the UMAP visualizations in \Fig{dca_umaps} and our knowledge of the physical environments sampled by each dataset.
For example, the DCA-computed precision scores validate our expectations that \texttt{Carbon\_GAP\_JCP2020} is the most diverse (highest row-average in \Fig{dca_prmat}) of the four datasets, and that \texttt{C\_npj2020} is well captured by most of the other datasets (highest column-average).
The precision scores also allow us to make additional useful observations, e.g, \texttt{C\_Gardner\_2020} is largely distinct (low precision \textit{and} low recall) from both \texttt{C\_npj2020} and \texttt{CA-9}, which is supported by the minimal overlap seen \Fig{dca_umaps}.

These types of insights can be extremely valuable to DDIP dataset developers when designing test sets or seeking to merge existing training sets to fit a more general model.
For example, when merging a new training set into an existing one, a low precision score indicates that the new data is introducing new information into the training set.
Similarly, a high recall score indicates that the new data may be over-sampling regions of configurational space that are already well-represented by the existing data, therefore leading to an effective increased weighting of those regions of space on the loss function which can affect model performance and training metrics.
Furthermore, precision and recall scores could help to identify more suitable test sets,
where it may be desirable that the test set have low precision and high recall (e.g., to detect possible overfitting), low precision and low recall (e.g., to test model generalizability/zero-shot capacity), or any range of values in between these limits depending upon the goal of the test.
Use of DCA, or related metrics, can provide a more systematic approach to dataset construction, which can help to address the known issues of high redundancy and correlation in DDIP training sets and materials data \cite{Li2023_redundancy,Heid2023,Vita2023}, and will likely be essential moving forward in the field to ensure that datasets are not inhibiting the ability of researchers to properly assess model generalizability.
We would like to emphasize that DCA is just \textit{one} example of a method which could lead to better dataset design -- other techniques (e.g., dataset roughness \cite{Aldeghi2022}, information imbalance \cite{Glielmo2022}, or entropy-based metrics \cite{Karabin2020}) may be equally valuable, and should be further developed alongside the ColabFit Exchange. 
Importantly, because the ColabFit Exchange houses an ever-growing number of diverse datasets, it can help facilitate large-scale benchmarking and analysis of new methods (such as DCA), and provide insights across many unique datasets.

\subsection{Example fitting workflow}

In order for the ColabFit Exchange to be usable in practice, it is important that the datasets be easily accessed and interacted with by a variety of DDIP fitting frameworks \cite{Bartok2010,Singraber2021,Batzner2021,Rohskopf2023}.
While this is achievable by writing simple I/O operations for exporting datasets from the ColabFit Exchange as extended XYZ files, then re-formatting to integrate with external software, a more streamlined approach would be one which operates directly on the native ColabFit Exchange data structures and ties in with necessary simulation and validation packages.
ColabFit Exchange datasets can be utilized for end-to-end DDIP development entirely within the KIM ecosystem, taking advantage of existing tools such as KLIFF~\cite{Wen2022kliff} for model training, and OpenKIM~\cite{Tadmor2011,Elliott2011,tep13,karls:bierbaum:2020} for model testing, archiving, and deployment.
As an example of such an end-to-end workflow, we use KLIFF to train a spline-based MEAM potential~\cite{Baskes1992,Lenosky2000} for lithium (Li) using the \texttt{mlearn-Li} training dataset, which has been used along with its other elemental counterparts for model benchmarking~\cite{Zuo2020,Vita2021}.
KLIFF supports seamless loading of ColabFit Exchange datasets, training of physics-based IPs and arbitrary machine learning DDIPs based on the PyTorch library \cite{pytorch}, and exporting of KIM-compliant models that can then be seamlessly deployed to a variety of molecular simulation packages that support the KIM standard including ASE~\cite{Larsen2017}, DL\_POLY~\cite{dlpoly}, GULP~\cite{gulp} and LAMMPS~\cite{Thompson2022} (see Ref.~\citenum{OKusing23} for a full list).

We fit the spline-based MEAM potential to energy and forces utilizing 7 knots per spline and an inner and outer cutoff radius of 2.4~\AA \space and 5.1~\AA, respectively. 
The model achieved training (testing) set energy and force RMSEs of 1.55 (1.65)~meV/atom and 0.049 (0.046)~eV/\AA, respectively. 
\begin{table}[!h]
    \centering
    \caption{Computed lattice constant ($a$), elastic constants ($c_{ij}$), bulk modulus ($K$), vacancy formation, migration and diffusion activation energies ($E_{\rm v}$, $E_{\rm m}$, $E_{\rm a}$), and surface energies ($E_{\rm s}$) of bcc Li using the spline-based MEAM potential and DFT. Relative errors between MEAM and DFT values are also shown. All values for the fitted MEAM potential were computed using the OpenKIM framework. DFT reference values are taken from Materials Project~\cite{Jain2013,deJong2015,Tran2016}  (mp-135) except for vacancy energies which are taken from Ref.~\citenum{Ko2017}.}
    \begin{center}
        \begin{tabular}{cccc|cccc}
            \hline\hline
            Property & MEAM &  DFT & Rel. Error & $E_{\rm s}$ (Jm$^{-2}$) & MEAM & DFT & Rel. Error\\
            \hline
            $a$ (\AA) & 3.44  & 3.44 & 0.000 & (100) & 0.466 & 0.462 & 0.009\\
            $c_{11}$ (GPa) & 17 & 15 & 0.133 & (110) & 0.448 & 0.501 & 0.106\\
            $c_{12}$ (GPa) & 13 & 13 & 0.000 & (111) & 0.516 & 0.544 & 0.051\\
            $c_{44}$ (GPa) & 10 & 11 & 0.091 & (210) & 0.473 & 0.506 & 0.065\\
            $K$ (GPa) & 14 & 14 & 0.000 & (211) & 0.505 & 0.538 & 0.061\\
            $E_{\rm v}$ (eV) & 0.455 & 0.481 & 0.054 &(310) & 0.473 & 0.497 & 0.048\\
            $E_{\rm m}$ (eV) & 0.055 & 0.042 & 0.309 & (311) & 0.494 & 0.527 & 0.063\\
            $E_{\rm a}$ (eV) & 0.510 & 0.523 & 0.025 &(320) & 0.603 & 0.504 & 0.196\\
            & & & & (321) & 0.499 & 0.534 & 0.065\\
            & & & & (322) & 0.510 & 0.535 & 0.047\\
            & & & & (331) & 0.489 & 0.521 & 0.061\\
            & & & & (332) & 0.592 & 0.524 & 0.130\\

            \hline\hline
        \end{tabular}
    \end{center}
    \label{tab:meam}
\end{table}
Additional material property predictions of the trained potential can be seen in Table~\ref{tab:meam}.
The potential performs well across all computed properties, with the largest relative errors being those of surface energy predictions (0.196 for the (320) surface).
This decreased performance of the model on surface energy predictions is not surprising, given the relatively small number of surface COs present in the training set (one CO per surface).
The one exception is the vacancy migration energy, $E_{\rm m}$, which has a higher relative error than the surface energies due its small magnitude.
These results, along with results from automated verification checks on model integrity can be viewed on \url{https://openkim.org/cite/MO_386038428339_000}  \cite{Fuemmeler2023}, where the model has been archived along with $>$600 other curated and contributed models for a wide variety of chemical and material systems. This potential can be invoked in a portable fashion~\cite{Afshar2023} within a variety of simulation platforms as explained above. We note that this example is only meant as a demonstration of how the interoperability ColabFit/KLIFF/OpenKIM leads to a streamlined fitting workflow. A potential major benefit of the ColabFit Exchange is the ability to leverage multiple datasets for DDIP development utilizing strategies such as transfer learning~\cite{Zaverkin2023} and meta-learning~\cite{Allen2023}. However, these approaches are still very much an open scientific question, which we will seek to address in future work pertaining to the ColabFit project.

\section{Contributing}

As with many open-source projects, the utility of the ColabFit Exchange will grow in proportion to the amount of engagement it receives from the research community.
Contributions from the community may come in many forms. To name just a few possibilities, this could include: developing and uploading new DDIP training sets; training models to existing datasets and documenting performance metrics; improving the metadata in the database by adding labels to COs or defining new, meaningful CSs; or developing new tools (like those discussed in \Sec{dca}) for characterizing dataset distributions.

Given that we foresee uploading training sets as being the most likely manner in which users will contribute to the ColabFit Exchange, we provide here some guidance on how users may best approach this task.
The simplest way to contribute is through the Github repository at \url{https://github.com/colabfit/data-lake}, where instructions are provided for uploading data or requesting that the ColabFit team obtain existing data from the literature.
Datasets contributed in this manner will be reviewed and parsed by the ColabFit team before submission to the database.
In order to streamline the process of constructing useful and interpretable datasets, the following best practices should be followed by researchers interested in uploading their data to the ColabFit Exchange:

\begin{itemize}
    \item DSs should be given meaningful, human-readable names. These need not be unique, since DSs are identified by their hashes, but it is useful if they are, in order to avoid confusion.
    \item Training/testing splits should be provided as separate DSs.
    \item DSs and CSs should be given concise descriptions outlining their contents. Discussions of the type of data contained within them (molecular, condensed matter, etc.) and their target applications (catalysis, radiation damage, drug discovery, benchmarking, etc.) are particularly useful.
    \item As much as possible, COs should be organized into conceptually meaningful CSs.
    \item As much as possible, COs should be given human-readable labels.
    \item All metadata required for reproducing a calculation (e.g., INCAR files) should be provided if possible.
    \item Computed properties should be adjusted to conform to existing PDs (a list of which can be found at \url{colabfit.org}). New PDs should be defined sparingly. Units must always be specified, when applicable.
\end{itemize}

Two of the most common, and challenging, issues that we struggled to overcome during the process of gathering datasets for the ColabFit Exchange were when dataset developers 1) used custom, poorly-documented storage formats for their data; or 2) did not define any conceptual groupings over their label which could be translated into CSs or CO labels.
In general, we recommend the use of the Extended XYZ format as commonly used by ASE \cite{Larsen2017}, and the application of at least rudimentary labels on COs (e.g., ``ground state", ``liquid", ``strained", etc.).
For examples of well-constructed datasets, we point the reader to Refs.~\citenum{Byggmstar2019}, \citenum{Bartk2018}, and \citenum{Wood2019}, whose authors we commend for publishing datasets with many desirable traits: 1) open-access, 2) well-documented storage formats, 3) good labeling of COs, and 4) clearly-defined groupings of COs.

While the Github repository is the simplest approach to contributing data, it relies upon a significant amount of effort from the ColabFit team in order to review and process the uploaded data, or to read through journal articles and contact authors to obtain access to their datasets.
As an alternative, for those users who are able and willing, the \texttt{colabfit-tools} package provides all of the necessary code to manually parse your dataset into the data objects described in \Sec{methods} (see \url{https://github.com/colabfit/colabfit-tools} for examples).
This takes a large burden off of the ColabFit team, and can greatly accelerate the upload process.

\section{Conclusion}\label{sec:conclusion}

In this work we have developed a flexible and robust data standard that we applied to atomistic property data to construct the ColabFit Exchange, the first database of its kind specializing in data for data-driven interatomic potential generation typically employing machine learning techniques.
At the time of writing (September 2023), the ColabFit Exchange contains 139 curated datasets and is actively being expanded, with particular emphasis on benchmarking datasets---those, which have been well tested, clearly documented, and shown to be suitable for analyzing aspects of model quality and guiding future development of reliable IPs.
Along with the development of the ColabFit Exchange, we demonstrated the usefulness of DCA for identifying and characterizing overlapping regions of datasets, which can help to further guide dataset generation towards populating under-sampled regions of configurational and compositional space, thus improving the generalizability of the resultant DDIPs.
Finally, we have shown how the data within the ColabFit Exchange can be utilized for end-to-end development of IPs within the KIM ecosystem, providing the benefits of seamless data retrieval, model exporting for use with major simulation software packages, and automated model verification, testing, and archiving on \url{https://openkim.org}.
While our current focus is on atomistic data, specifically properties commonly applied to IP development, our framework is flexible enough to support a variety of different data ``silos'', e.g., databases for meta-materials, bio-sequences, etc., which may become another application of the project in later work.

Future efforts of the ColabFit project will be to explore additional techniques for analyzing novel properties of datasets, like those described in \Sec{dca}, which have been shown in some cases to correlate with generalizability and fitting errors of resultant models, and to develop metrics based on precision and recall scores for characterizing the utility of test sets.
Further code development will also be done in order to expand the \texttt{colabfit-tools} package, with a focus on developing a Python API for accessing/contributing data, constructing datasets, and running consistency checks over contributed data (which is currently only done by hand).
Perhaps most important for leveraging ColabFit's full potential will be gaining a better understanding of data interoperability and novel training strategies that can incorporate data across multiple datasets, levels of theory, and simulation parameters. As the ColabFit Exchange grows and matures, we anticipate it being an important tool for developing novel (meta-)learning strategies, which have recently been applied to atomistic datasets with promising results~\cite{Allen2023}.

We invite the community to upload data via the Github repository at \url{https://github.com/colabfit/data-lake} and will work closely with dataset developers who wish for their data (and models) to be findable, accessible, interoperable, and reusable.

\section*{Data and Code Availability}

The entirety of the ColabFit Exchange can be found at \url{https://colabfit.org}. The \texttt{colabfit-tools} package can be found at \url{https://github.com/colabfit/colabfit-tools}.

\section*{Acknowledgements}

This research was supported through the National Science Foundation (NSF) under grant OAC-2039575. S.M. acknowledges the Simons Center for Computational Physical Chemistry for financial support. This work was supported in part through the NYU IT High Performance Computing resources, services, and staff expertise.
J.A.V. acknowledges the DIGI-MAT program at UIUC, which is supported by the National Science Foundation under Grant No. 1922758.
The authors wish to acknowledge the Minnesota Supercomputing Institute (MSI) at the University of Minnesota for providing resources that contributed to the results reported in this paper.

\subsection*{Author Contributions}
\noindent
Conceptualization: EBT, SM, RSE, JAV\\
Methodology: EBT, SM, RSE, JAV, EGF\\
Software: JAV, EGF, AG\\
Validation: EBT, SM, JAV, EGF\\
Investigation: EBT, SM, JAV, EGF\\
Data Curation: JAV, EGF, AQT, GPW, AG\\
Writing - Original Draft: JAV, EGF, GPW\\
Writing - Review \& Editing: EBT, SM, JAV, EGF, GPW\\
Visualization: JAV, EGF, GPW\\
Supervision: EBT, SM

\section*{Competing interests}

The authors declare that they have no competing interests.

\bibliographystyle{unsrt}  
\bibliography{bibliography}

\appendix

\clearpage

\begin{center}
    \fontsize{16}{16}\selectfont Supporting Information for\\[12pt]
    \fontsize{14}{14}\selectfont\textbf{ColabFit Exchange: open-access datasets for data-driven interatomic potentials}\\[12pt]
    \fontsize{12}{12}\selectfont Joshua A. Vita$^{1}$, Eric G. Fuemmeler$^{2}$, Amit Gupta$^{2}$, Gregory P. Wolfe$^{3}$, Alexander Quanming Tao$^{2}$, Ryan S. Elliott$^{2}$, Stefano Martiniani$^{3,4,5}$, and Ellad B. Tadmor$^{2,\ast}$\\[12pt]
    \fontsize{11}{11}\selectfont $^1$Department of Materials Science and Engineering, University of Illinois at Urbana-Champaign, Urbana, IL 61801, USA \\
    \fontsize{11}{11}\selectfont $^2$Department of Aerospace Engineering and Mechanics, University of Minnesota, Minneapolis, MN 55455, USA \\
    \fontsize{11}{11}\selectfont $^3$Center for Soft Matter Research, Department of Physics, New York University, New York, NY 10012, USA \\
    \fontsize{11}{11}\selectfont $^4$Simons Center for Computational Physical Chemistry, New York University, New York, NY 10012, USA \\
    \fontsize{11}{11}\selectfont $^5$Courant Institute of Mathematical Sciences, New York University, New York, NY 10012, USA \\
    \fontsize{11}{11}\selectfont $^\ast$Corresponding author: tadmor@umn.edu
\end{center}

\renewcommand{\thefigure}{S\arabic{figure}}
\setcounter{figure}{0}
\renewcommand{\thetable}{S\arabic{table}}
\setcounter{table}{0}
\setcounter{page}{1}

\section{Additional DCA details}

We provide here further details regarding the DCA algorithm in order to help readers better understand the method and interpret the corresponding results in this work.
As is mentioned in the main text, DCA is a technique which was originally intended to be used for comparing a single dataset under multiple representations, but which we have instead applied in this work to comparing multiple datasets under the same representation.
The DCA algorithm can be roughly broken down into three key steps:
\begin{enumerate}
    \item \textbf{Manifold approximation}, where Voronoi cells and the corresponding Delaunay graph are approximated in the original representation space. In short, this is done by first constructing the Voronoi cells using Monte Carlo sampling (as outlined in the previous work of the DCA authors \cite{Polianskii2019}), then by projecting rays originating from each data point to build the Delaunay graph edges, connecting points with adjacent Voronoi cells.
    \item \textbf{Component distillation}, where the Delaunay graph is clustered into connected components using the HDBSCAN \cite{McInnes2017} algorithm.
    \item \textbf{Component evaluation}, where consistency and quality scores (\Eqn{dca_cq}) are computed for each connected component in order to identify ``fundamental components" which have both high consistency and high quality.
\end{enumerate}
The fundamental components are then used to compute the global precision and recall scores, as defined in \Eqn{dca_pr}.

As outlined in Ref.~\citenum{Poklukar2022}, DCA has five tunable hyper-parameters:
\begin{itemize}
    \item $T$, the number of rays sampled from each data point for finding points with adjoining Voronoi cells in order to build the Delaunay graph edges. Larger values will incur higher memory and computation costs, but are more likely to find all of the correct edges in the graph.
    \item $B$, the ``sphere coverage", which is used as a threshold value for filtering edges in the Delaunay graph to reduce memory consumption. A value of 1.0 means that no filtering is performed.
    \item $mcs$, the minimum cluster size used by the HDBSCAN algorithm when clustering points during sub-graph construction.
    \item $t_c$, the consistency threshold used for defining a fundamental component.
    \item $t_q$, the quality threshold used for defining a fundamental component.
\end{itemize}
In our experiments we found that the default values for some of these hyper-parameters ($B=1.0$, $mcs=10$) were reasonable choices, where the DCA results did not change significantly when varying the values.
We chose to decrease $T$ from the default value of $10^4$ down to $10^2$, which greatly improved the speed of the calculations without altering the results.
This is consistent with the observations from Ref.~\citenum{Poklukar2022}, where they show that the DCA results that were relatively consistent across multiple choices of $T$, $B$, and $mcs$.

\begin{figure}[!ht]
    \centering
    \includegraphics[scale=0.75]{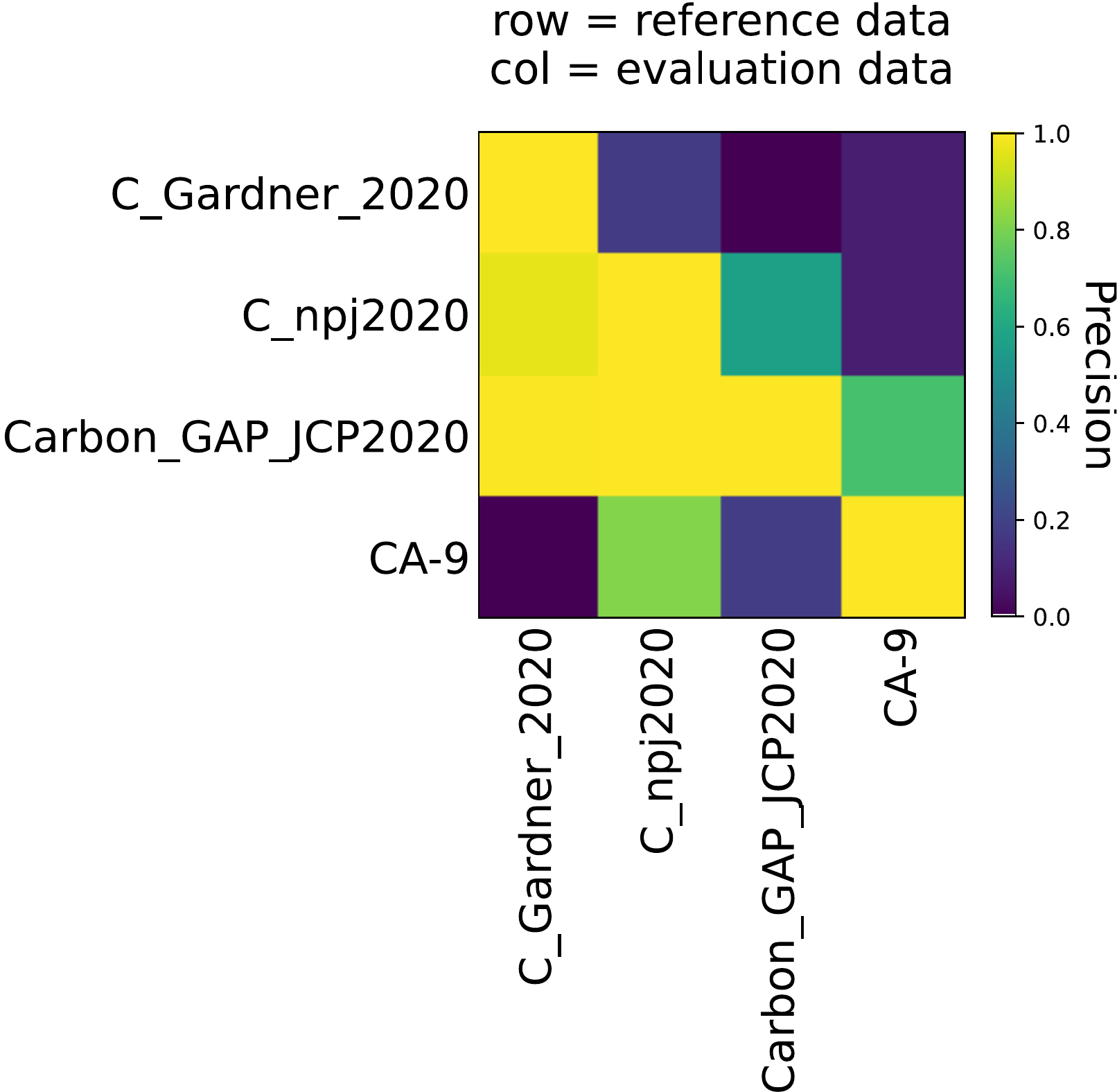}
    \caption{A version of Fig. 5 computed using the default DCA values of $t_c = t_q =0$. Note that the \texttt{Carbon\_GAP\_JCP2020}/\texttt{C\_Gardner\_2020} cell has a precision score of 1, which contradicts our expectations based on the qualitative analysis from \Fig{dca_umaps}.}
    \label{fig:dca_prmat_zero_thresh}
\end{figure}

\begin{figure}[!ht]
    \centering
    \includegraphics[width=\linewidth]{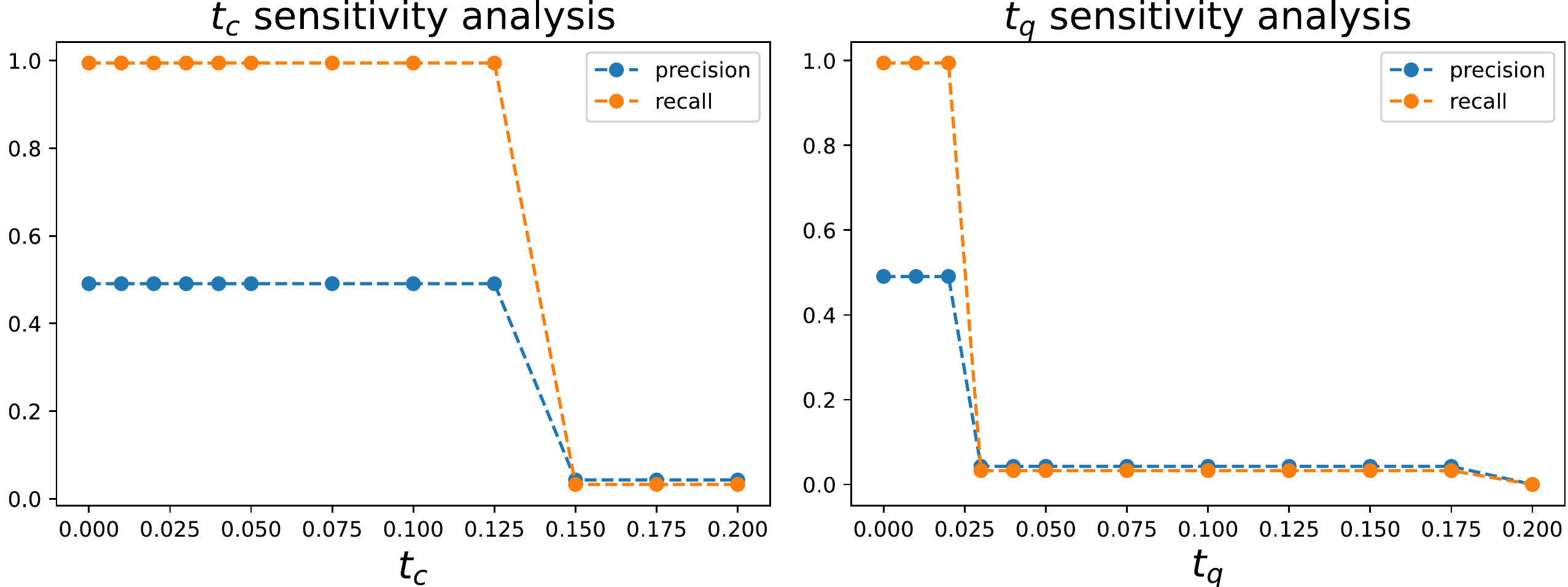}
    \caption{Precision and recall scores for different (left) consistency thresholds, $t_c$, and (right) quality thresholds, $t_q$, for the \texttt{C\_Gardner} and \texttt{C\_GAP} datasets. For the $t_c$ ($t_q$) sensitivity analysis the $t_q$ ($t_c$) threshold is held constant at 0.}
    \label{fig:dca_cq_sens}
\end{figure}

The consistency and quality thresholds, $t_c$ and $t_q$, however, can greatly alter the DCA results when choosing values that are too small.
This can result in the identification of fundamental components which do not adequately reflect the distributions of the reference and example datasets within the components.
For example, we observe in \Fig{dca_prmat_zero_thresh} that the default DCA values can result in erroneously high precision scores in some cases (e.g., when a component has just a single point from the example set in it).
Given the poor behavior observed in \Fig{dca_prmat_zero_thresh}, we performed a more thorough sensitivity analysis of the precision and recall scores with respect to the threshold values $t_c$ and $t_q$.
The results of this sensitivity analysis are shown in \Fig{dca_cq_sens}, where it can be seen that there are two cases where poor choices of $t_c$ or $t_q$ can yield unexpected results: 1) when the threshold values are too small, such that a component with even a small number of poorly-mixed points is identified as being ``fundamental", and 2) when the threshold values are too large, such that even reasonably well-mixed components are not considered to be fundamental, and the precision and recall scores are then computed as 0 based on \Eqn{dca_pr}.
Given the behavior observed in \Fig{dca_cq_sens}, we chose to use threshold values of $t_c=t_q=0.01$ in this work.
However, we emphasize that similar sensitivity analysis should be performed when using DCA with new datasets.

\end{document}